\documentclass[reprint, amsmath,amssymb, aps, pra, floatfix]{revtex4-1}

\usepackage{graphicx}
\usepackage{dcolumn}
\usepackage{bm}

\usepackage{units}
\usepackage[squaren]{SIunits}
\usepackage{amsmath,amsfonts,amssymb}

\allowdisplaybreaks[2]

\newcommand{\im}{\ensuremath{\mathrm{i}}}

\begin{document}

\title{\underline{Pha}se-sensitive \underline{n}uclear \underline{t}arget \underline{s}petroscop\underline{y} (PHANTASY)}

\author{Benedikt \surname{Herkommer}}

\author{J\"org \surname{Evers}}

\affiliation{Max Planck Institute for Nuclear Physics, Saupfercheckweg 1, 69117 Heidelberg, Germany}

\date{\today}

\begin{abstract}
M\"ossbauer nuclei feature exceptionally narrow resonances at hard x-ray energies, which render them ideal probes for structure and dynamics in condensed-matter systems, and a promising platform for x-ray quantum optics and fundamental tests. However, a direct spectroscopy at modern x-ray sources such as synchrotrons or x-ray free electron lasers is challenging, because of the broad spectral bandwidth of the delivered x-ray pulses, and because of a limited spectral resolution offered by x-ray optics and detectors.
To overcome these challenges, here, we propose a spectroscopy technique based on a spectrally narrow reference absorber that is rapidly oscillating along the propagation direction of the x-ray light. The motion induces sidebands to the response of the absorber, which we scan across the spectrum of the unknown target to gain spectral information. The oscillation further introduces a dependence of the detected light on the  motional phase at the time of x-ray excitation as an additional controllable degree of freedom. We show how a Fourier analysis with respect to this phase enables one to selectively extract parts of the recorded intensity after the actual experiment, throughout the data analysis. This allows one to improve the spectral recovery by removing unwanted signal contributions. Our method is capable of gaining spectral information from the entire measured intensity, and not only from the intensity at late times after the excitation, such that a significantly higher part of the signal photons can be used. Furthermore,  it not only enables one to measure the amplitude of the spectral response, but also its phase.
\end{abstract}

\maketitle

.
\section{Introduction}

Spectroscopy is an indispensable tool to study matter and its dynamics. Starting from initial work in the visible regime, by now its different variants are established across vast ranges of the electromagnetic spectrum, covering many orders of magnitude on the frequency scale, and there is a continuous progress in further advancing the different spectroscopy techniques. Over the last years, in particular spectroscopy at soft and hard x-ray energies has undergone a revolutionary development, due to improved x-ray sources and advances in x-ray optical elements and detection techniques~\cite{
doi:10.1146/annurev-physchem-032511-143720,RevModPhys.88.015007,Kraus2018,springerreview,shvydko,optics2,3rdGenSyncrotronReview,adams_scientific_2019}. Towards hard x-ray energies, however, atomic resonances are broadened due to their intrinsically low lifetime, such that it is difficult to find sharp electronic resonances. An interesting alternative are M\"ossbauer nuclei, which feature spectrally narrow transition in the hard x-ray regime~\cite{Mossbauer1958,ralf,moessbauer_story_book}. These are narrow due to the M\"ossbauer effect, i.e., the recoil-free interaction between  x-rays and matter~\cite{Mossbauer1958}. They  can be viewed as almost ideal two-level systems, have very high quality factor, and form the basis for a wide range of applications~\cite{moessbauer_story_book,doi:10.1002/9781118714614,Oshtrakh2019,ralf,doi:10.1080/09500340.2012.752113,Ruffer2014,Rohlsberger2014}. However, due to the exceptionally small line width, dispersive or diffractive elements to directly spectrally resolve their response are not readily available.

\begin{figure*}[t]
	\includegraphics[width=\textwidth]{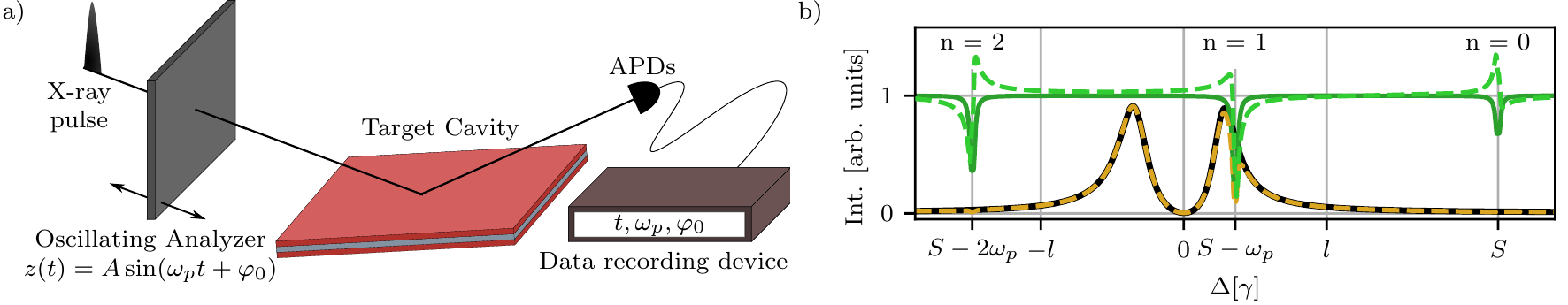}
	\caption{(Color online) (a) Schematic setup of the PHANTASY spectroscopy scheme. The goal is to measure the spectral response of an unknown target containing M\"ossbauer nuclei using a short, spectrally broad x-ray pulse delivered by an accelerator-based source. The target is illustrated by a thin-film cavity containing the resonant nuclei. To introduce an energy  selectivity, a spectrally narrow reference absorber also containing nuclei is introduced into the x-ray pathway. This reference absorber performs rapid oscillations in direction along the beam propagation. Subsequently,  the scattered radiation is measured using avalanche photo diodes in dependency of the time $t$ after excitation, the oscillation frequency of the piezo $\omega_p$, the oscillatory offset $\varphi_0$ of the piezo, and the oscillation amplitude $p$.
	In (b), the operation principle of the spectroscopy approach is  illustrated. The target spectrum to be measured is shown as the black solid line with asymptotic value 0. The oscillations introduce sidebands to the spectrum of the reference absorber shown as green solid and dashed lines with asymptotic values 1. The reference absorber has the central frequency $S$. One of the sidebands ($n=1$) at $S-\omega_p$ is scanned across the target spectrum by varying the oscillation frequency $\omega_p$. 	The dashed orange dashed line shows the product of the analyzer spectrum and the target energy spectrum, which determines the signal measured by the detector in the time domain. Importantly, the spectrum of the reference absorber depends on the phase of the motion $\varphi_0$ at the time of excitaiton, indicated by the shape of the sidebands for $\varphi_0=0$ (solid) and $\varphi_0=\nicefrac{\pi}{2}$ (dashed line). We show that this dependence can be used to remove all unwanted contributions to the detection signal, and to recover amplitude and phase of the target response.
    }
	\label{fig:4_Principle_of_Phantasy}
\end{figure*}
 
One way of performing spectroscopy on M\"ossbauer nuclei is to use radioactive x-ray sources, which offer sufficiently narrow source line widths to probe the spectra of unknown samples~\cite{moessbauer_story_book}. In contrast, modern pulsed x-ray sources are orders of magnitude broader than typical nuclear resonances even after monochromatization using crystal optics~\cite{ralf}. One approach to overcome this challenge is to measure in the time domain~\cite{ruby,PhysRevLett.54.835}, which allows one to separate the off-resonant background and the scattered signal light via temporal gating.
 However, this approach does not directly provide spectral information, since detectors only register the intensity without phase information in a limited time after the excitation, such that a direct Fourier transformation of the response is not possible. Another ansatz is to use nuclei themselves to monochromatize the incoming x-ray light for the subsequent experiment, which enables the direct measurement of spectra in the energy domain~\cite{PhysRevB.41.9545,SynchrotronMoessebauerSource,SynchrotronMoessebauerSource2,SynchrotronMoessebauerSource3,PhysRevLett.102.217602,masuda_development_2008}.

A qualitatively different approach is to add a spectrally narrow reference absorber to the experiment, and to scan its resonance across the spectrum of the unknown sample, e.g., via Doppler shifts induced by relative motions between sample and absorber with constant velocity~\cite{PhysRevB.54.16003,Coussement2000,PhysRevB.61.4181,LateTimeIntegrationKilian,LTI-LambShift-Roehlberger,XrayPhaseDetection,Haber2016}. 
The combined response of unknown sample and reference absorber integrated over time then enables one to recover the energy spectrum of the  sample, as function of the detuning of the reference absorber. However,  it was found that  the recovered spectra depend sensitively on the integration range~\cite{Odeurs1998,LTI-LambShift-Roehlberger,ralf}, e.g., because of Fourier time-window effects introduced by gating away the off-resonant photons in the exciting x-ray pulse. One way of addressing this issue is stroboscopic detecion~\cite{PhysRevB.65.180404,PhysRevB.67.104423}, in which the scattered light is measured in periodically spaced short intervals only. In recent experiments, instead, an integrating only over detection times late after the initial excitation was used~\cite{LTI-LambShift-Roehlberger,Rohlsberger2012,LateTimeIntegrationKilian,XrayPhaseDetection,Haber2016}, which we denote as Doppler-drive method in the following.  This approach can be understood by noting that the desired spectral information is encoded in those signal photons which interacted with both,  the sample and  the reference absorber, and therefore were delayed twice until they reach the detector, thereby contributing to the late detection times. In contrast, at early detection times, the desired spectroscopy information is masked by those  photons which interacted with either the target or the analyzer, but not both, and therefore were delayed only once.  As a result, the requirement to restrict the analysis to late detection times severely restricts the part of the signal photons contributing to the recovery of the spectrum due to the near-exponential decay of the signal.
Furthermore, this method as well as most other techniques for M\"ossbauer nuclei only allow one to measure the magnitude of the target response, but not its phase. However, the stroboscopic detection can access phase information~\cite{PhysRevB.67.104423}, and there are interferometric techniques~\cite{PhaseDetectionSturhahn,XrayPhaseDetection}.

Here, we put forward a spectroscopy method to characterize the amplitude and phase of the response of an unknown sample containing M\"ossbauer nuclei, which we denote as phase-sensitive nuclear target spectroscopy (PHANTASY). It also uses a spectrally narrow reference absorber, but employs oscillatory motions of the reference absorber, instead of the conventional motion with constant velocity, see Fig.~\ref{fig:4_Principle_of_Phantasy}(a). This motion gives rise to a sequence of sidebands to the absorber resonance, which we scan across the unknown target response [see Fig.~\ref{fig:4_Principle_of_Phantasy}(b)]. The key difference to the previous approaches is that the oscillatory motion introduces a sensitivity of the detection signal to the phase $\varphi_0$ of the absorber motion at the time of arrival of the x-ray pulse. We exploit this by establishing the Fourier analysis with respect to $\varphi_0$ as a powerful analysis tool. It enables one to split the experimentally accessible detection signal into different components after the actual experiment, throughout the data analysis. Most notably, this technique enables one to remove all unwanted contributions to the detection signal arising from photons which only interacted with the target, but not with the reference absorber. As a consequence, the integration range to recover the target spectrum can significantly be enlarged towards shorter detection time, thereby considerably increasing the fraction of signal photons contributing to the recovery of the signal. To facilitate the spectral recovery, we derive a compact expression for the detection intensity using a Fourier analysis with respect to $\varphi_0$ and an additional spectral filtering, which can both be applied in the data analysis. Interestingly, because of the sensitivity to $\varphi_0$, this expression provides access not only to the amplitude, but also to the phase of the desired target response. Next to oscillatory motions, we also introduce a second motional pattern, which is more challenging to implement in practice, but directly enables one to generate a tunable single-line absorber rather than a sequence of sidebands. In this case, no additional filtering is required, and the measurement also has the favorable dependence on $\varphi_0$.

The required oscillatory or step-like motions of the reference absorber can be realized, e.g., using piezo transducers, and are well-established in the M\"ossbauer community~\cite{SuperEarlyOscillationPaper, VibrationWithSidebands,CalibrationOfMBSpectrometerViaOscillation,RadiativeCouplingAndDecouplingByMotion,NuclearPertubationByUltrasound,UltrasoundVibrationUsingSychrotronRadiation,RadiationBurstOlga,SciencePaper_PiezoPhase,Olga_Oscillation_Paper,HELISTO1981177,Olga_Oscillation_Paper,coherentcontrol}. They have been used, e.g., to calibrate M\"oessbauer spectrometer \cite{CalibrationOfMBSpectrometerViaOscillation}, to dynamically couple and decouple different nuclear targets~\cite{RadiativeCouplingAndDecouplingByMotion}, to explore the propagation of the x-ray through nuclear absorbers~\cite{NuclearPertubationByUltrasound,UltrasoundVibrationUsingSychrotronRadiation}, to shape given x-ray pulses favorably in the time- or energy domain~\cite{HELISTO1981177,RadiationBurstOlga,SciencePaper_PiezoPhase,Olga_Oscillation_Paper}, or for the coherent control of nuclear dynamics using x-ray light~\cite{coherentcontrol}. In particular, it is known that 
oscillatory motions give rise to spectral sidebands~\cite{HELISTO1981177,Olga_Oscillation_Paper}, and experimental data has already been selectively analyzed as function of the motional phase $\varphi_0$ at the time of arrival of the x-ray pulse~\cite{Olga_Oscillation_Paper}. However, we are not aware of a separation of the data in terms of a Fourier analysis with respect to $\varphi_0$, nor the spectroscopy  applications introduced below.

The article is organized as follows. In Sec.~\ref{chap:2_Theory}, we introduce the theoretical background of nuclear resonance scattering, and the time- and energy representations of the response of stationary and moving targets containing M\"ossbauer nuclei to incident x-ray radiation. Further, we introduce the motional patterns used in our analysis. The following Sec.~\ref{sec:phantasy} then introduces the PHANTASY  method. Starting from the combined response of moving absorber and unknown target, we in particular establish the Fourier-analysis with respect to the phase $\varphi_0$ as a powerful tool in Sec.~\ref{sec:phifilter}. Next, we numerically explore the capabilities of the spectroscopy method in Sec.~\ref{chap:Evaluation}, and compare it to the existing Doppler-shift-based spectroscopy method. Finally, Sec.~\ref{sec:summary} discusses and summarizes our results.

\section{Theoretical background \label{chap:2_Theory}}

\subsection{General setting}

The general setup of our method is depicted in Fig.~\ref{fig:4_Principle_of_Phantasy}(a). The aim is to measure the spectral response of an unknown target  using a short x-ray pulse delivered by an accelerator-based source such as a synchrotron radiation facility. In the figure, the target is illustrated by a thin-film cavity containing resonant nuclei. Due to their short temporal duration, the spectrum of the x-ray pulses is orders of magnitude broader than the target spectrum to be measured, such that a scan of the source cannot give spectral information on the target. Also the avalanche photo detector cannot resolve the desired spectral information.

To overcome this challenge, a spectrally narrow single-line reference absorber is introduced into the x-ray beam. Different from previous spectroscopy approaches, this reference absorber is mounted on a piezo transducer which rapidly oscillates the position of the analyzer parallel to the beam propagation direction. Further, the intensity of the scattered light is measured as a function of time $t$ after the x-ray excitation, the oscillation frequency of the piezo $\omega_p$, the oscillation phase $\varphi_0$ at the time of excitation, and as function of the oscillation amplitude $A$. To this end, we propose to use an event-based detection system, which can record different quantities for each signal photon separately~\cite{PhysRevLett.104.087601,SciencePaper_PiezoPhase,coherentcontrol} for the later data analysis.

The operation principle of our spectroscopy technique is illustrated in Fig.~\ref{fig:4_Principle_of_Phantasy}(b). The rapid oscillation introduces sidebands in the spectral response of the reference absorber, which are separated by the oscillation frequency $\omega_p$. As example, the transmittance of the oscillating absorber is shown as the green solid line in the figure, with the fundamental frequency and two sidebands. This transmittance depends on the phase $\varphi_0$, which determines the spectral shape of the different sidebands. To obtain spectral information, the first sideband of the analyzer is scanned across the unknown spectrum of the target via a suitable change of the oscillation frequency $\omega_p$. To allow for a selective coupling of one sideband to the target, an energy offset $S$ between the resonance frequency of the stationary analyzer  and the target spectrum is introduced, such that the other sidebands are sufficiently far away from the target spectrum during the scan. This can be achieved, e.g., by mounting the target on a Doppler drive moving with constant velocity, or by adding a linear component to the oscillatory analyzer motion.

In the following, we revisit the theoretical background to describe this setting. Afterwards, our spectroscopy approach will be discussed in detail in Sec.~\ref{sec:phantasy}.

\subsection{\label{sec:stationary}Response of the stationary nuclear reference analyzer}
We start by analyzing the propagation of x-rays through a stationary analyzer featuring a single nuclear resonance. This is the starting point for the reference absorber used to probe the spectra of the unknown sample. In the frequency domain,  the spectral response of the analyzer can be written as~\cite{ralf,2Level_Energy_Domain}
\begin{align}
\hat{R}(\omega) = \exp\left(\frac{-\im b}{\omega - \omega_0 + \frac{i\gamma}{2}}\right)\label{eq:2_SingleLineResonanceAbsorptionEnergyDomain},
\end{align}
where $\gamma$ is the single-nucleus linewidth and $\omega_0$ the resonance frequency of the nuclei. Here and in the following, a ``hat'' indicates that the formula refers to the frequency domain. The parameter
\begin{align}
b =  \frac{\pi \rho_N f_\mathrm{LM} \gamma  d}{k_0^2 (\alpha + 1)}
\end{align}
describes the thickness $d$ of the analyzer foil, and is proportional to its optical depth. Here, $\rho_N$ is the nuclear number density, $k_0$ the wave number of the resonance, $f_\mathrm{LM}$ the Lamb-M\"ossbauer factor, and $\alpha$  the internal conversion factor.
In case of the archetype M\"ossbauer nucleus $^{57}$Fe, the numerical values of these parameters are $\rho_N = \unit{83.18}{nm^{-3}}$,  $k_0 = \unit{73.039}{{nm}^{-1}}$, $f_\mathrm{LM} \approx 0.8$,  $\alpha = 8.56$, $\gamma = \unit{4.7}{neV}$  and $\omega_0 = \unit{14.4}{keV}$.

With the help of Eq.~(\ref{eq:2_SingleLineResonanceAbsorptionEnergyDomain}), the outgoing x-ray field $\hat{E}_\mathrm{out}(\omega)$ can be calculated from the incident field $\hat{E}_\mathrm{in}(\omega)$ as
\begin{align}
\hat{E}_\mathrm{out}(\omega) = \hat{R}(\omega) \: \hat{E}_\mathrm{in}(\omega).
\end{align}
If the incident field is spectrally much broader than the nuclear resonance, as it is the case for temporally short synchrotron pulses, the input field has no relevant frequency dependence $E_\mathrm{in}(\omega) \approx 1$. Consequently, the outgoing field is proportional to the response function.

For the following analysis, it is convenient to rewrite  Eq.~(\ref{eq:2_SingleLineResonanceAbsorptionEnergyDomain}) as
\begin{align}
	\hat{R}(\omega) &= 1 - \hat{R}_\mathrm{S}(\omega)\,,\\
	\hat{R}_\mathrm{S}(\omega) &=  1 - \exp\left(\frac{-\im b}{\omega - \omega_0 + \frac{i\gamma}{2}}\right) \,.
\end{align}
Here, the ``1'' correspond to the zeroth scattering order and describes light which passes the analyzer without interaction with the nuclei. The remaining part $\hat{R}_\mathrm{S}(\omega)$ contains all higher order of the scattering between x-rays and nuclei, and the subscript $\mathrm{S}$ indicates this part of the response due to scattering. For a thin analyzer, an expansion of $\hat{R}_\mathrm{S}(\omega)$ to leading order in the thickness parameter $b$ leads to a Lorentzian frequency response, as expected for a single resonance. However, higher-order scattering processes in thicker analyzers lead to non-Lorentzian spectral responses.

The corresponding response function in the time-domain can be obtained by Fourier-transform as~\cite{ralf}
\begin{subequations} \label{eq:2_SingleLineResponse_TimeDomain}
\begin{align}
	R(t) &= %
\frac{1}{2\pi} \int_{-\infty}^{\infty} \hat{R}(\omega)  e^{-\im \omega t} d\omega \notag \\[2ex]
&= \delta(t) - R_\mathrm{S}(t)\,,\\
R_\mathrm{S}(t) &=    \theta(t)\, \sqrt{\frac{b}{t}} \: J_1\left (2\sqrt{bt}\right)\:  e^{-\im \omega_0 t}\: e^{- \frac{\gamma}{2}t} \,.
\end{align}
\end{subequations}
In the time domain, the zeroth-order response corresponding to x-ray  passing through the analyzer without interactions is given by the $\delta(t)$-contribution, whereas the contributions $R_\mathrm{S}(t)$ arising from the interaction between x-rays and nuclei are delayed to times $t\geq0$ due to the narrow nuclear line width.

To calculate the outgoing field amplitude $E_\mathrm{out}(t)$ in the time-domain, the input field $E_\mathrm{out}(t)$ is convoluted with the response function,
\begin{align}
	E_\mathrm{out}(t) = R(t) \ast E_\mathrm{in}(t)\,.
\end{align}

\subsection{Response of the moving\\nuclear reference analyzer \label{sec:2_PhaseManipulationViaMotion} }
Next, we consider modifications to the nuclear response discussed in Sec.~\ref{sec:stationary} due to fast mechanical  motions of the absorber. To this end, we transform the incoming pulse from the laboratory frame into the moving rest frame of the analyzer. Within this frame, the interaction between x-rays and analyzer can be calculated as in the static case. Afterwards, the scattered light is transformed back to the laboratory frame.  For a near-instantaneous $\delta(t)$-like incident x-ray pulse as it is provided by a synchrotron radiation source, one finds for the response of the moving analyzer~\cite{SciencePaper_PiezoPhase,olga_PRA_PAPER,Olga_Oscillation_Paper}
\begin{align}
R_\mathrm{moving}(t) = e^{\im k_0 \left[z(t) - z(0)\right]}\: R(t), \label{eq:2_Response of moving target in lab frame}
\end{align}
where $z(t)$ describes the motion of the target.

\subsubsection{Harmonic oscillation}
An important class of analyzer motions are harmonic oscillations~\cite{Olga_Oscillation_Paper,UltrasoundVibrationUsingSychrotronRadiation,UltrasoundOscillation,SuperEarlyOscillationPaper, VibrationWithSidebands}, for which the piezo motion can be described by
\begin{align}
	z(t) =    A\: \sin(\omega_pt + \varphi_0) \label{eq:2_Osscillating piezo function}\,.
\end{align}
Here, $A$ is the oscillation amplitude,  and $\varphi_0$ describes the phase of the oscillation at the time $t=0$ at which the incident x-ray pulse hits the analyzer. Using Eq.~\eqref{eq:2_Response of moving target in lab frame}, we find for the response of an oscillating single-line analyzer  described by Eq.~\eqref{eq:2_SingleLineResponse_TimeDomain}
\begin{align}
R_\mathrm{osc}(t) = \delta(t)  - e^{\im p \left[\sin(\omega_p t + \varphi_0) - \sin(\varphi_0) \right]} R_\mathrm{S}(t) \,,
\label{eq:2_Oscillating Single line, time domain, compact}
\end{align}
where $p = A \cdot k_0$ is a dimensionless quantity characterizing the  oscillation amplitude.

To transform this equation to the frequency domain, we expand the exponential function using the Jakobi-Anger relationship,
\begin{align}
e^{\im p \sin(x + \varphi_0)} = \sum_{n = -\infty}^{\infty} J_n(p) e^{\im n(x + \varphi_0)}\,, \label{eq:2_Jakobi-Anger Relation}
\end{align}
and find
 \begin{align}
	\hat{R}_\mathrm{osc}(\omega) &= 1 - \sum_{n = -\infty}^{\infty} \alpha_n \: \hat{R}_\mathrm{S}(\omega + n\omega_p) \,, \\\label{eq:2_EnergySpectrumOscillatingSL}
      \alpha_n &= J_n(p) \: e^{\im n  \varphi_0 - \im p \sin(\varphi_0)}\,.
\end{align}
We thus recover the well-known result that the oscillation gives rise to spectral sidebands, which are separated by multiples of the oscillation frequency $\omega_p$. Each sideband comprises the stationary response $\hat{R}_\mathrm{S}$, shifted to the corresponding center frequency $\omega_0 - n \omega_p$, and multiplied by the prefactors $\alpha_n$.
These prefactors not only determine the relative weights of the different sprectral sidebands via the Bessel functions $J_n(p)$, but also contain relative phases depending on $\varphi_0$, which modify the spectral shape of the sidebands from Lorentzian to more general Fano line shapes. An example for this is given in Fig.~\ref{fig:4_Principle_of_Phantasy}, where as an example the sidebands $n=0,1,2$ are shown for $\varphi_0 = 0$ (solid green line) and $\varphi_0 = \nicefrac{\pi}{2}$ (dashed green line).

For applications in spectroscopy, we will find in the following discussion that it is favorable to realize situations in which a single sideband is of relevance. In this case, approximating to the $n=1$ sideband only, Eq.~\eqref{eq:2_EnergySpectrumOscillatingSL} reduces to
\begin{align}\label{single-line}
\hat{R}_\mathrm{osc,SL}(\omega) = 1 - J_1(p)\: e^{\im \varphi_0 - \im p \sin(\varphi_0)} \:\hat{R}_\mathrm{S}(\omega +  \omega_p)\,.
\end{align}

\subsubsection{\label{single-absorber}Phase-sensitive single line absorber}
In the case of an oscillating analyzer, the single-line case can only be realized in an approximate way, or using further processing of the experimental data, as discussed in Sec.~\ref{sec:phantasy}. However, there is an alternative motion, which directly leads to a single spectral line, while preserving a variable phase $\varphi_0$ in the spectral response.  It is given by
\begin{align}
	z(t) = z(0) + \frac{1}{k_0}\bigl[\theta(t) \varphi_0 +  \omega_p  t   \bigr] \,,
\end{align}
which corresponds to a step-like jump immediately after the excitation due to the incident short x-ray pulse in order to impose the relative phase $\varphi_0$, and  a linear motion to induce a Doppler shift by $ \omega_p$, thereby tuning the resonance energy of the single spectral line. In the frequency domain, the corresponding response is
\begin{align}
\hat{R}_\mathrm{PSSL} = 1 - e^{i \varphi_0} \: \hat{R}_\mathrm{S}(\omega + \omega_p)\,.
\end{align}
This result qualitatively agrees to the single line approximation of the harmonic motion Eq.~(\ref{single-line}), except for a different form of the prefactor, which is not of relevance for the spectroscopy applications discussed in Sec.~\ref{sec:phantasy}.

This motional pattern is ideally suited for applications in spectroscopy, and does not require removing the contributions of additional sidebands from the experimental data in a postprocessing step. However, it is considerably more challenging to precisely implement and control the step-like motion experimentally, because of the requirement to abruptly displace the analyzer immediately after the excitation via the x-ray pulse. We note, however, that only the relative motion between analyzer and target is of relevance. For this reason, it is possible to split the motion across the two. For example, one could apply the linear motion to the analyzer using a conventional Doppler-drive, and a variable step-like jump immediately after the excitation to the target to introduce the phase sensitivity, as realized in~\cite{SciencePaper_PiezoPhase}.

\section{\label{sec:phantasy}Phase sensitive nuclear target specroscopy (PHANTASY)}

With the theoretical background of Sec.~\ref{chap:2_Theory} at hand, we now proceed by discussing the phase-sensitive nuclear target spectroscopy method. We consider an unknown target containing nuclei, which is exemplified by the case of an x-ray cavity containing nuclei in Fig.~\ref{fig:4_Principle_of_Phantasy}. The goal is to determine the spectral response $\hat{R}_\mathrm{T}(\omega)$ of this target, or equivalently the temporal response function $R_\mathrm{T}(t)$. 
For this, we insert the harmonically oscillating analyzer foil in front of the target, such that the outgoing x-ray field at the detector is given by
\begin{align}
\hat{E}_\mathrm{D}(\omega) &=    \hat{R}_\mathrm{T}(\omega) \: \hat{R}_\text{osc}(\omega) \: \hat{E}_\mathrm{in}(\omega) \notag  \\
&=  \hat{R}_\mathrm{T}(\omega) - \nonumber \\
&\qquad \sum_{n = -\infty}^{\infty} \alpha_n \hat{R}_\mathrm{T}(\omega) \hat{R}_\mathrm{S}(\omega - S + n \omega_p)\,.
\label{eq:4_Electric Field at detector, exact}
\end{align}
Here, we again neglect $\hat{E}_\mathrm{in}(\omega) \approx$~const. for simplicity, assuming the case of spectrally broad synchrotron radiation. We further introduce a frequency offset $S$ between the center frequency of the target's response and that of the analyzer's response, which will become important in the later analysis.

To fully exploit the power of the phase-sensitive spectroscopy, we assume an event-based detection, in which for each signal photon the time of arrival $t$, the phase $\varphi_0$ at the time of excitation, the oscillation frequency $\omega_p$ and the oscillation amplitude $p$ are recorded, similar to that employed in~\cite{SciencePaper_PiezoPhase,coherentcontrol}. One approach is to record a trace of the time-dependent voltage applied to the transducer inducing the motion of the analyzer around each photon arrival time. Fitting a sine function to this trace provides both, $\varphi_0$ and $p$.

In order to relate the experimental data recorded in the time domain to Eq.~(\ref{eq:4_Electric Field at detector, exact}), one would also like to transform the latter to the time domain,
\begin{align}
\label{eq:problem}
 E_{\mathrm{D}}(t) = \frac{1}{2\pi}\int_{-\infty}^{\infty} d\omega \: \hat{E}_\mathrm{D}(\omega) \: e^{-i\omega t}\,.
\end{align}
However, this is not possible in a straightforward way, due to the unknown frequency dependence of the target response. The experimental data in turn cannot directly be transformed in to the frequency domain, since only the light's amplitude is recorded by the detector, but not its phase.  In the following, we will show how one can overcome this challenge, and thereby extract the amplitude and the phase of the target response from the experimentally accessible data.

\subsection{Sensing Head Approximation \label{sec:4_SensingHeadApproximation}}

As a first step, we employ an approximation which allows us to analytically perform the transformation Eq.~(\ref{eq:problem}) to the time domain. This approximation is also employed in the Doppler-drive spectroscopy method~\cite{LTI-LambShift-Roehlberger,PhD_Kilian,LateTimeIntegrationKilian}, and we denote it as ``sensing head approximation'' in the following, illustrating the role of the reference absorber. We start by rewriting Eq.~(\ref{eq:problem}) using Eqs.~(\ref{eq:4_Electric Field at detector, exact}) and (\ref{eq:2_SingleLineResponse_TimeDomain}) as
\begin{align}
\label{eq:approx1}
E_{\mathrm{D}}(t) &=
{R}_\mathrm{T}(t)  - \sum_{n = -\infty}^{\infty}
\frac{\alpha_n }{2\pi} \times \nonumber \\
&\quad \times\int_{-\infty}^{\infty} d\omega \: e^{-i\omega t}\: \hat{R}_\mathrm{T}(\omega) \hat{R}_\mathrm{S}(\omega - S + n \omega_p)\,.
\end{align}
The key idea of the approximation now is to assume that the spectral response of the analyzer is narrow as compared to that of the target, which in practice is realized by using thin analyzer foils. This assumption allows us to expand the frequency-dependence of the target response in the last line of Eq.~(\ref{eq:approx1}) around the respective sideband frequencies of the oscillating analyzer,
\begin{align}
 &\hat{R}_\mathrm{T}(\omega) \approx \hat{R}_\mathrm{T}(\omega_0 + S -  n\omega_p) e^{\im \tau (\omega - \omega_0 - S + n \omega_p)}\,,
 \label{eq:4_Fundametal approximation}
\end{align}
where
\begin{align}
\tau_n =  \left. \frac{\partial \mathrm{arg} [\hat{R}_\mathrm{T}(\omega)]  }{\partial \omega} \right|_{\omega = \omega_0 + S - n \omega_p} .\label{eq:4_Phase_approximation_higher_order}
\end{align}
Inserting this approximation into Eq.~\eqref{eq:approx1}, it is now possible to perform the  Fourier transformation, to give
\begin{align}
&E_D (t) \approx     R_\mathrm{T}(t) +  \sum_{n = -\infty}^{\infty} \alpha_n \: e^{-\im( S - n\omega_p)t}\times \nonumber \\
 &\qquad \quad \  \  \times \hat{R}_\mathrm{T}(\omega_0 + S - n \omega_p) \: R_\mathrm{S}(t - \tau_n) \:e^{-\im \omega_0 \tau_n}\,.
\label{eq:4_E at detector, approximated, time domain}
\end{align}
As a result of this approximation, we find that the time-dependent signal $E_D (t)$ at the detector contains information about the desired frequency-response of the target $\hat{R}_\mathrm{T}(\omega_0 + S - n \omega_p)$.

From Eq.~(\ref{eq:4_E at detector, approximated, time domain}), we evaluate the intensity registered by the detector as
\begin{subequations}
\label{eq:4_AbsSquare-atDetektor-exact}
\begin{align}
&I_\mathrm{D} = |E_\mathrm{D}(t)|^2 =   I_\mathrm{T} + I_\mathrm{Sq} + I_\mathrm{Re}\,, \\[2ex]
& I_\mathrm{T} =  |R_\mathrm{T}(t)|^2\,,\\
&I_\mathrm{Sq} = \left| \sum_{n = -\infty}^{\infty} R_\mathrm{S}(t-\tau_n) e^{-\im \omega_0 \tau_n }	\alpha_n \right .\nonumber \\
& \left . \qquad  \ \times  \hat{R}_\mathrm{T}(\omega_0 + S - n \omega_p)	\: e^{\im n \omega_p t}\right|^2 \,,\\[2ex]
&I_\mathrm{Re} = 2 \:\mathrm{Re} \left[ R^*_\mathrm{T}(t) \sum_{n = -\infty}^{\infty} \alpha_n\: e^{-\im (S - n \omega_p)t}\right. \nonumber\\
&\left . \qquad  \ \hat{R}_\mathrm{T}(\omega_0 + S - n \omega_p)\: R_\mathrm{S}(t - \tau_n) e^{-\im  \omega_0 \tau_n} \right]\,.
\end{align}
\end{subequations}
The first addend depends only on the time-dependent response of the target. The second part $I_\mathrm{Sq}$ is the sum term in Eq.~(\ref{eq:4_AbsSquare-atDetektor-exact}) squared. The final part $I_\mathrm{Re}$ is the interference contribution, which will turn out to be most useful for applications in spectroscopy.

It is apparent from Eq.~(\ref{eq:4_AbsSquare-atDetektor-exact})  that the detection signal is a complicated mixture of contributions arising from different sidebands in the spectrum of the oscillating absorber. One way of simplifying this result is to replace the oscillating absorber by the phase-sensitive single-line absorber introduced in Sec.~\ref{single-absorber}. In this case, all sums over $n$ in Eq.~(\ref{eq:4_AbsSquare-atDetektor-exact}) disappear, substantially simplifying the expression. Alternatively, we will show in the following how a comparable single-line information can be gained from the total detection signal in case of an oscillating absorber, by employing two different filter methods.

\subsection{\label{sec:phifilter}Disentangling the detector signal\\using the $\varphi_0$ - filter}

Next, we introduce a method to exploit the dependence of the detection signal on $\varphi_0$ to selectively extract certain parts from it. In an experiment, the harmonic oscillation typically is not frequency-locked to the repetition rate of the x-ray pulses, such that throughout the measurement  the phase $\varphi_0$ is automatically sampled over its entire range $[0,2\pi[$. An event-based detection scheme registering this phase for each signal photon separately then provides the basis for a powerful analysis technique introduced below.

The key idea of the $\varphi_0$-filter  is to perform a Fourier analysis of the detection signal in Eq.~\eqref{eq:4_AbsSquare-atDetektor-exact} with respect to $\varphi_0$, in order to separate the total detection signal into the various frequency components of this phase. In the following, the Fourier-conjugate variable to $\varphi_0$ will be called $f$, such that
\begin{align}
 I_{x} = \sum_{f=-\infty}^{\infty} \: I_{x}^f\: e^{-if\varphi_0}\,,
\end{align}
for $x\in\{\mathrm{T}, \mathrm{Sq}, \mathrm{Re}\}$. Hence, $I_x^f$ is the Fourier coefficient describing the signal part oscillating with the frequency of $f\varphi_0$.

The first addend $I_\mathrm{T}$ of Eq.~\eqref{eq:4_AbsSquare-atDetektor-exact}  has no dependency on $\varphi_0$, such that
\begin{align}
I_\mathrm{T} = I_\mathrm{T}^{f=0}\,.
\end{align}
Since the other contributions only depend on $\varphi_0$ via $\alpha_n$, we next evaluate the  Fourier-transformations of the combinations of $\alpha_n$ which appear in Eq.~\eqref{eq:4_AbsSquare-atDetektor-exact}.
For $I_\mathrm{Sq}$, the absolute square leads to contributions of the form  $\alpha_n \alpha_m^*$, which transform to
\begin{align}
\mathcal{F}( \alpha_n \alpha^*_m; \varphi_0, f) = 2 \pi J_n(p) J_m(p) \:  \delta(f+ n - m )\,. \label{eq:4_Fourier transform of alpha *m alpha n}
\end{align}
$I_\mathrm{Re}$ contains the $\varphi_0$-dependent contributions $\alpha_n $ and $\alpha_n^*$. Using again the Jakobi-Anger relation Eq.~\eqref{eq:2_Jakobi-Anger Relation}, we obtain
\begin{subequations}
\label{eq:4_Fourier transform of alpha n}
 \begin{align}
\mathcal{F}(\alpha_n; \varphi_0, f) &= 2 \pi \: J_n(p)  \sum_{m = -\infty}^{\infty} J_m(p) \delta(f + n - m)\,,\\
\mathcal{F}(\alpha_n^*; \varphi_0, f) &= 2 \pi \: J_n(p)  \sum_{m = -\infty}^{\infty} J_m(p) \delta(f + m - n)\,.
\end{align}
\end{subequations}
Using these results, we now can evaluate the $f$-component of detection signal, to obtain
\begin{subequations}
\label{eq:afterphifilter}
\begin{align}
I_\mathrm{T}^f =& |R_\mathrm{T}(t)|^2 \: \delta_{f,0}\,,\\[2ex]
I^f_\mathrm{Sq}(t) =&  \sum_{n = -\infty}^{\infty}
\biggl[%
 J_n(p) J_{f + n}(p) e^{-\im f \omega_pt} e^{-i\omega_0(\tau_n - \tau_{n+f})}\nonumber \\
&\times %
R_\mathrm{S}(t - \tau_n) \: R^*_\mathrm{S}(t - \tau_{f+n})
\nonumber \\
&\times\hat{R}_\mathrm{T}(S - n \omega_p) \:\hat{R}^*_\mathrm{T}\bigl(S - (n+f)\omega_p\bigr)  \biggr]\,,
%
%
\label{eq:4_E_Sq_spectral} \\[2ex]
I^f_\mathrm{Re}(t) = &  \sum_{n = -\infty}^{\infty}
\biggl[ J_n(p) J_{n+f}(p) \: R^*_\mathrm{T}(t)\nonumber \\
& \times \hat{R}_\mathrm{T}(\omega_0 + S - n \omega_p)\: e^{-\im(  S - n \omega_p)t}  \nonumber \\
& \times R_\mathrm{S}(t - \tau_n)\: e^{-\im \omega_0 \tau_n}   +  c.c. \biggr]\,.
\label{eq:4_E_Re_spectral}
\end{align}
\end{subequations}
As a result, we find that the Fourier analysis with respect to $\varphi_0$ separates the total detection signal into different components. Since the Fourier transform only has contributions at integer values of $f$, this separation can reliably be performed even in the presence of experimental imperfections. Afterwards, the different components can selectively be analyzed or combined. Most notably, choosing $f\neq 0$ enables one to completely remove the contribution $I_{\textrm{T}}$ which is favorable since it does not contain the desired dependence on $\hat{R}_{\mathrm{T}}$, and usually  renders the desired components in the time-dependent detection signal inaccessible at early detection times.

\subsection{Disentangling the detector signal\\using the $t$ - filter \label{sec:4_Fourier-tFilter-theory}}

\begin{figure}[t]
	\includegraphics[width = \columnwidth]{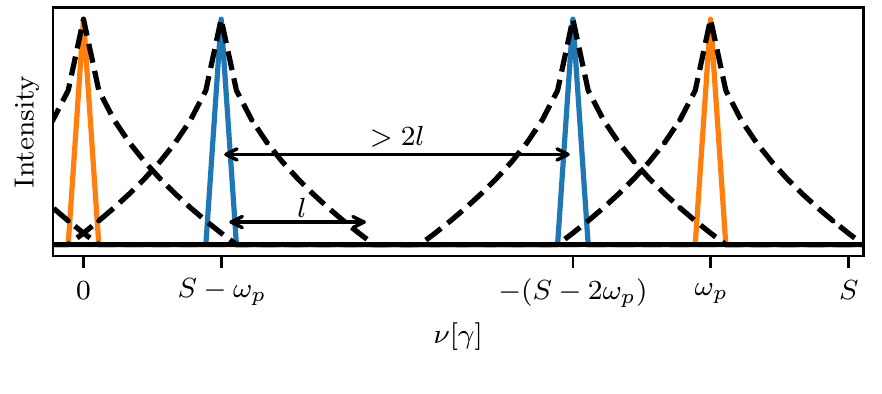}
	\caption{(Color online) Schematic representation of the spectral sideband positions contributing to the detection signal, as function of the frequency $\nu$. Each sideband is indicated as a triangle. Positions of the sidebands in Eq.~\eqref{eq:4_E_Sq_spectral} are shown in orange at $0, \omega$, whereas those of \eqref{eq:4_E_Re_spectral} are indicated in blue. The broadening of the peaks due to the convolution with the unknown target spectrum is indicated by the black dashed lines. $l$ denotes the half-width of this broadening. To ensure that there is no spectral overlap between the broadened peaks such that a $t$-filter can be applied, the distances between the relevant peaks should be separated at least by $2l$.}
	\label{fig:4_peakPositioninNuSpace}
\end{figure}

The $t$-filter discussed in this section is analogous to the $\varphi_0$-filter, but exploits Fourier transformations between the time- and frequency spaces. In the following, the conjugated variable to $t$ in the Fourier space is called $\nu$. While the $\varphi_0$-filter conveniently allows one to remove the  contribution $I_{\textrm{T}}$ from the detected signal, the remaining signal contributions in Eqs.~(\ref{eq:afterphifilter}) still comprise sums over different sidebands $n$ in each $f$-component. The purpose of the $t$-filter therefore is to further select a single sideband component from the  $\varphi_0$-filtered signal. However, the $t$-filter is more difficult to implement than the $\varphi_0$-filter, since spectral components contributing to the detection signal are not discrete, in contrast to the integer $f$-components contributing to the $\varphi_0$-dependence. The origin of this is that the different sideband contributions are broadened in $\nu$-space due to the convolution with the target response, such that they may spectrally overlap. This is illustrated in Fig.~\ref{fig:4_peakPositioninNuSpace}, where the positions of the different sidebands are symbolized using triangles, whereas their broadening due to the convolution with the target response is indicated by the dashed black lines.

Because  the sidebands separation depends on $\omega_p$, in order to implement the $t$-filter, we demand that this frequency is large enough such that the different spectral components created by the harmonic motion do not spectrally overlap. To quantify this criterion, we define $l$ such that the relevant frequency range of the target spectrum is contained within the interval $[\omega_0 - l, \omega_0+l]$. In Fig.~\ref{fig:4_peakPositioninNuSpace}, $l$ corresponds to the half width of the dashed broadened lines.

We start by extracting the frequencies of the dominant time dependencies of the sideband components from Eqs.~(\ref{eq:afterphifilter}), given by $\pm (S-n\omega_p)$ for $I^{f}_\mathrm{Re}$ and by $\pm f \omega_p$ for $I^{f}_\mathrm{Sq}$. The $\nu$-range shown in Fig.~\ref{fig:4_peakPositioninNuSpace} includes two sidebands of $I^{f}_\mathrm{Re}$ at $S-\omega_p$ and $-(S-2\omega_p$) (blue triangles), as well as two sidebands of $I^{f}_\mathrm{Sq}$ at $0$ and $\omega_p$ (orange triangles).

Since in the proposed setup the sideband with $n = 1$ is used to scan the target resonance, the aim of the $t$-filter is to extract the  component with dominant frequency $\pm (S- 1 \cdot \omega_p)$ from the measured signal. The  dominant frequencies of the other spectral components therefore should be separated by more than $2l$ in the $\nu$-space.
To quantify this condition, we find from Fig.~\ref{fig:4_peakPositioninNuSpace} that the closest unwanted frequency component of $I^{f}_\mathrm{Re}$ to $(S-\omega_p)$ is $-(S-2\omega_p)$. Their frequency difference is $|3\omega_p - 2S|$. Since $\omega_p$ has to be scanned within the interval  $[S-l,S+l]$ to cover the entire target spectrum, the frequency difference is smallest for $\omega_p = S-l$. In this case, the frequency difference evaluates to $|S-l3|$, which should be larger than $2l$. As a result, we find the condition
\begin{align}
 S > 5l \label{eq:conds}
\end{align}
in order to be able to successfully apply the $t$-filter. This condition can be interpreted in a straightforward way. With increasing $S$, the oscillation frequency $\omega_p$ also has to increase in order to scan across the target spectrum. Due to this increase, the different sidebands move further apart. This leads to a minimum value for $S$ in order to be able to separate the spectral components, on a scale set by the width of the target spectrum $2l$.

Next, we analyze $I^{f}_\mathrm{Sq}$. In the above case, all components with dominant evolution frequency $\pm f \omega_p$  for $f\neq 0$ are separated from the desired frequency component $\pm (S-\omega_p)$ by more than $2l$.
The $f=0$ contribution of $I^{f}_\mathrm{Sq}$, however, overlaps with the $\pm (S-\omega_p)$  component of $I^{f}_\mathrm{Re}$ for $\omega_p \approx S$, i.e., in the center of the target spectrum, such that the two cannot be separated by the $t$-filter alone.

As a result, we thus find that  the $t$-filter also is a powerful method to extract particular information from the measured data, if condition Eq.~(\ref{eq:conds}) is fulfilled. However, like the $\varphi_0$-filter, the $t$-filter alone cannot be used to extract the response of a single sideband component of the data. Therefore, a combination of both is needed to achieve the reduction to a single sideband component.

\subsection{Extracting the contribution of a single sideband  by combining the $\varphi_0$- and the $t$-filters}

Here, we show that the combination of the $\varphi_0$- and the $t$-filter enables one to extract the response of a single spectral component from the measured data. To this end, we first use the $\varphi_0$-filter to select the $f = \pm 1$ components from the data. The relevant contributions of the  signal evaluate to
\begin{subequations}
\begin{align}\label{eq:fpm1}
 &I^{f=1}_\mathrm{T}(t)e^{-i\varphi_0} + I^{f=-1}_\mathrm{T}(t)e^{i\varphi_0} = 0\,,\\[2ex]
&I^{f=1}_\mathrm{Sq}(t)e^{-i\varphi_0} +  I^{f=-1}_\mathrm{Sq}(t)e^{i\varphi_0}   \nonumber \\
& \qquad \qquad = (\dots) e^{i\omega_p t} + (\dots) e^{-i\omega_p t} \,,
 \\[2ex]
 &I^{f=1}_\mathrm{Re}(t)e^{-i\varphi_0} + I^{f=-1}_\mathrm{Re}(t)e^{i\varphi_0}\nonumber \\
 &=2\,\textrm{Re}\biggl\{\sum_{n = -\infty}^{\infty}
 J_n(p) \left[J_{n+1}(p)e^{-i\varphi_0}+ J_{n-1}(p)e^{i\varphi_0}\right] \:\nonumber \\
& \qquad \times  R^*_\mathrm{T}(t)\: \hat{R}_\mathrm{T}(\omega_0 + S - n \omega_p)\: e^{-\im(  S - n \omega_p)t}  \nonumber \\
& \qquad \times R_\mathrm{S}(t - \tau_n)\: e^{-\im \omega_0 \tau_n}  \biggr\}\,.
\end{align}
\end{subequations}
Here, we for clarity only indicate the residual dominant time dependency in the $I^{\pm 1}_\mathrm{Sq}$ contributions.
Note that this step removed the problematic $f=0$ component of $I_\mathrm{Sq}$ at $\nu = 0$ from the data. Therefore, now the $t$-filter can next be applied in order to select the dominant $(S - \omega_p)$ component. The filtered detection signal becomes
\begin{align}
\bar{I}_\mathrm{D} &= 2\,\textrm{Re}\biggl\{
 J_1(p)\left[J_{2}(p)e^{-i\varphi_0}+ J_{0}(p)e^{i\varphi_0}\right] \:\nonumber \\
& \qquad \times  R^*_\mathrm{T}(t)\: \hat{R}_\mathrm{T}(\omega_0 + S -   \omega_p)\: e^{-\im(  S -   \omega_p)t}  \nonumber \\
& \qquad \times R_\mathrm{S}(t - \tau_1)\: e^{-\im \omega_0 \tau_1}  \biggr\}\,. \label{eq:4_Equation_before_choice_of_p}
\end{align}
A straightforward comparison shows that Eq.~(\ref{eq:4_Equation_before_choice_of_p}) coincides with the result one would obtain by calculating the total detector signal within the single-line approximation of the oscillating analyzer in Eq.~(\ref{single-line}).
We therefore conclude that the $\varphi_0$- and the $t$-filter together enable one to extract the response of a single sideband from the full measured data.  Inspecting Eq.~(\ref{eq:4_Equation_before_choice_of_p}) further, we find that it contains the phase of the target response, even though the detector registers intensities only. This dependence arises due to the $\varphi_0$-dependence of the oscillating analyzer response, and offers the possibility to measure amplitude and phase of the desired target response, as we will demonstrate in the next Section.

A further simplification is achieved if we assume that the oscillatory motion has a particular modulation depth $p_0$  such that $J_0(p_0) = 0$. In this case,
\begin{align}
\bar{I}_\mathrm{D} &= 2\,\textrm{Re}\biggl\{
 J_1(p_0)J_{2}(p_0)e^{-i\varphi_0} \:\nonumber \\
& \qquad \qquad \times  R^*_\mathrm{T}(t)\: \hat{R}_\mathrm{T}(\omega_0 + S -   \omega_p)\: e^{-\im(  S -   \omega_p)t}  \nonumber \\
& \qquad \qquad \times R_\mathrm{S}(t - \tau_1)\: e^{-\im \omega_0 \tau_1}  \biggr\}\,. \label{eq:4_Equation_after_choice_of_p}
\end{align}

\subsection{Reconstruction of amplitude and phase\\of the target's spectral response}
With the analytical expression for the filtered detection signal Eq.~(\ref{eq:4_Equation_before_choice_of_p}) at hand, we are now in the position to discuss the recovery of the desired amplitude and phase of the  target spectrum. For simplicity, we illustrate the method for the case of Eq.~(\ref{eq:4_Equation_after_choice_of_p}), assuming a modulation depth such that $J_0(p_0) = 0$.
For this, we rewrite
\begin{subequations}
\begin{align}
  R_{\textrm{T}}(t) &= |R_{\textrm{T}}(t)|\: e^{i\varphi_{\textrm{T}}(t)}\,, \\
  \hat{R}_{\textrm{T}}(\omega) &= |\hat{R}_{\textrm{T}}(\omega)|\: e^{i\hat{\varphi}_{\textrm{T}}(\omega)}\,,
\end{align}
\end{subequations}
and make use of the relation $R_{\textrm{S}}(t) = |R_{\textrm{S}}(t)|\,\exp(-i\omega_0 t)$ [see Eq.~(\ref{eq:2_SingleLineResponse_TimeDomain})] to obtain
\begin{align}
 \label{eq:4_FinalResultFullSystem}
\bar{I}_{\textrm{D}} &=B(t, \tau_1) \cdot C(\omega_p) \cdot \cos[\varphi_0 + a(\omega_p, t)]\,,
\end{align}
where
\begin{subequations}
\begin{align}
B(t, \tau_1 ) &= 2 \,J_1(p_0) J_2(p_0) \, |R_\mathrm{S}(t - \tau_1)|\cdot |R_\mathrm{T}(t)| \,, \\[2ex]
C(\omega_p) &= |\hat{R}_\mathrm{T}(\omega_0 + S - \omega_p)| \,, \\[2ex]
a(t,\omega_p) &=  \hat{\varphi}_{\textrm{T}}(\omega_0 + S - \omega_p) - \varphi_{\textrm{T}}(t) \nonumber\\
&\qquad  - (\omega_0 + S-\omega_p)t \,.
\end{align}
\end{subequations}
This result can be interpreted in the following way. The filtered detection signal is proportional to the amplitude of the desired target spectrum $C(\omega_p)$. The corresponding phase $\hat{\varphi}_{\textrm{T}}$ is contained in the total phase $a$, and because of the dependence of the argument of the $\cos$ function on $\varphi_0$, a tomographic reconstruction of $\hat{\varphi}_{\textrm{T}}$ becomes possible by suitable variation of $\varphi_0$. Additionally, the total signal has a time-dependent prefactor $B(t, \tau_1)$, due to the individual decay of the responses of target and analyzer with increasing time. Note that Eq.~(\ref{eq:2_SingleLineResponse_TimeDomain}) mixes the temporal representation  $R_{\textrm{T}}(t)$ and its frequency version $ \hat{R}_{\textrm{T}}(\omega)$ in a single expression due to the sensing hat approximation.

\subsubsection{Fit of the experimental data}
\begin{figure}[t]
	\includegraphics{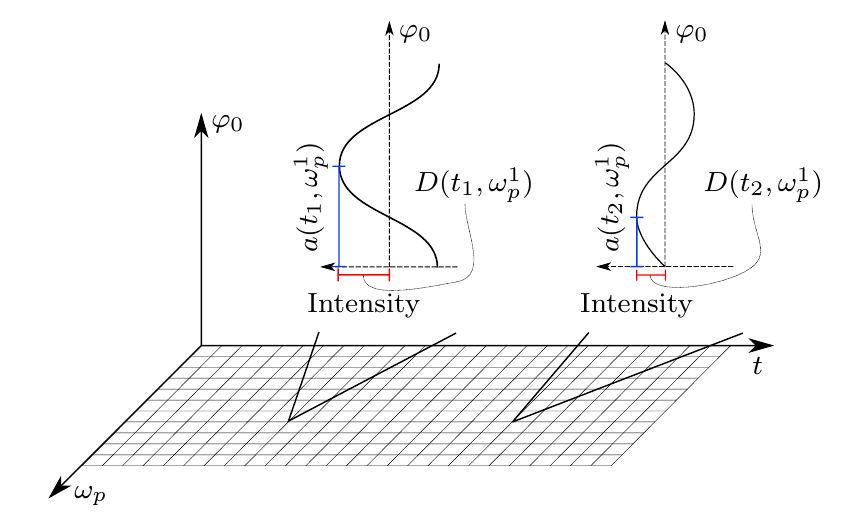}
	\caption{(Color online) Schematic illustration of the multidimensional data grid recorded in the experiment. For each pair ($\omega_p$, $t$), the filtered  intensity Eq.~(\ref{eq:4_FinalResultFullSystem}) has an oscillatory dependence in $\varphi_0$ direction. This cosine-dependence is captured via a fit as function of $\varphi_0$, yielding the amplitude $D(t, \omega_p)$ and the phase offset $a(t,\omega_p)$ in Eq.~(\ref{eq:fitmodel}). These allow one to reconstruct amplitude and phase of the target's spectral response.}
	\label{fig:4_3D-Grid with fit}
\end{figure}

As the first step of the recovery, the function
\begin{align}\label{eq:fitmodel}
 g(\varphi_0) = D \cos(\varphi_0  + a)
\end{align}
is fitted to the experimental data for each parameter pair $(\omega_p, t)$, see Fig.~\ref{fig:4_3D-Grid with fit}. Note that the possibility to perform this analysis crucially relies on the event-based multidimensional detection, which enables one to evaluate the experimental data with respect to any combination of the different variables.

\subsubsection{Reconstruction of the amplitude $|\hat{R}_{\textrm{T}}(\omega)|$\\of the target's spectral response}
For the reconstruction of the amplitude of the target spectrum, we use the fit parameter $D(t,\omega_p) = B(t,\tau_1 ) C(\omega_p)$. Like in the Doppler-drive spectroscopy method, we integrate this parameter over the time $t$, in order to make use of all recorded signal photons for the recovery. Note, however, that the $\varphi_0$- and the $t$-filter already removed all unwanted contributions from the detection signal. As a result, {\it a priori} the integration time range is not restricted to late times, as it is the case in the Doppler-drive spectroscopy method.

In addition to the increased measurement statistics, the time integration serves two further purposes in our method. First, it reduces the residual dependence on the parameter $\tau_1$ introduced because of the sensing-head approximation. $\tau_1$ depends on $\omega_p$ and therefore could lead to distortions of the recovered spectrum. However, it merely acts as a shift of $D$ in time. Therefore, its effect can be reduced using an integration over a larger time range. Second, our numerical simulations suggest that higher-order contributions not accounted for in the sensing-head approximation lead to additional oscillations in $D$ with time. Also these oscillations can be reduced using a time integration.
However, since the amplitude of the measured data decreases exponentially in time, the latter two effects are most effective if the contributions at all times contribute equally. To achieve this, the maximum of $D$ is normalized to unity at each instance in time $t$ separately, before the integration over all times $t$ is performed. Furthermore, since $D$ is proportional to the magnitude of the energy spectrum, the data is squared before the average is taken in order to obtain the spectral intensity.

\begin{figure}[t]
	\includegraphics{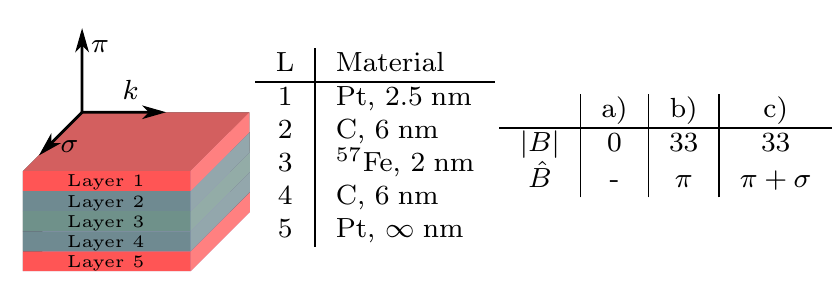}
	\caption{(Color online) Thin-film cavity structure used in the numerical analysis. $L$ labels the different layers of the cavity. Target spectra of varying complexity are achieved by considering three configuration for the magnetic field experienced by the nuclei in the target, as summarized in the table. The strength of the field $|B|$ is given in Tesla, and $\hat{B}$ indicates the orientation of the field.  The x-rays propagate in $k$-direction. }
	\label{fig:5_CavityMagnetisation}
\end{figure}

\subsubsection{Reconstruction of the phase $\hat{\varphi}_{\textrm{T}}(\omega)$\\of the target's spectral response}
For the reconstruction of the spectral phase, the fit parameter $a(t,\omega_p) = \hat{\varphi}_{\textrm{T}}(\omega_0 + S - \omega_p) - \varphi_{\textrm{T}}(t)   - (\omega_0 + S-\omega_p)t$  is used.
Due to the periodicity of the cosine function, the value of $a(t,\omega_p)$ is only determined modulo $2\pi$, or $\pi$ if the sign of $D$ is variable. This leads to a degeneracy of the fit parameters which needs to be corrected for.  Next, we subtract the $(\omega_0 + S - \omega_p)t$ contribution from $a(t,\omega_p)$. The remaining part only depends via the desired spectral phase $\hat{\varphi}_{\textrm{T}}$ on the oscillation frequency $\omega_p$, while the other parts only form an irrelevant overall phase offset. Finally, taking the modulus of $2\pi$ of the remaining part yields the desired spectral phase, up to an irrelevant overall phase.

Note that also for the recovery of the phase, it is useful to average the measured data over time. Since the parameter $a(t,\omega_p)$ does not depend on the amplitude $D(t,\omega_p)$, no normalization is necessary in this averaging.

\section{Evaluation Of the Phantasy Method \label{chap:Evaluation}}

\subsection{Numerical simulation}
In this Section, we numerically explore the potential of the PHANTASY method. As target, we consider a thin film cavity structure shown in Fig.~\ref{fig:5_CavityMagnetisation}, probed by the x-rays in grazing incidence. It comprises two layers of Pt acting as mirrors for the x-rays, and a guiding layer made of C. This guiding part further contains a thin layer of $^{57}$Fe featuring the resonances to be probed. In order to compare results for spectra of different complexity, we consider three different magnetic field configurations, summarized in Fig.~\ref{fig:5_CavityMagnetisation}. The case without magnetic splitting can be realized, e.g., by implementing the nuclear layer in the form of stainless steel~\cite{ralf}. In case of $\alpha$-Fe, a Zeeman splitting occurs, and the nuclei in general feature six different transition frequencies in the spectrum~\cite{ralf}. By applying weak external magnetic fields of different orientations, the magnetization axis of the nuclei can be aligned, thereby determining the coupling of the different transition dipole moments to the linearly polarized incident x-rays. In all cases, the cavity is irradiated by $\pi$-polarized x-ray light, and only $\pi$-polarized scattered light is measured.
For the analyzer, we consider a stainless steel foil  ($^{57}$Fe${}_{55}$Cr${}_{25}$Ni${}_{20}$)  of thickness of $d = \unit{1}{\mu m}$ that features a single absorption line.

\begin{figure}[t]
	\includegraphics{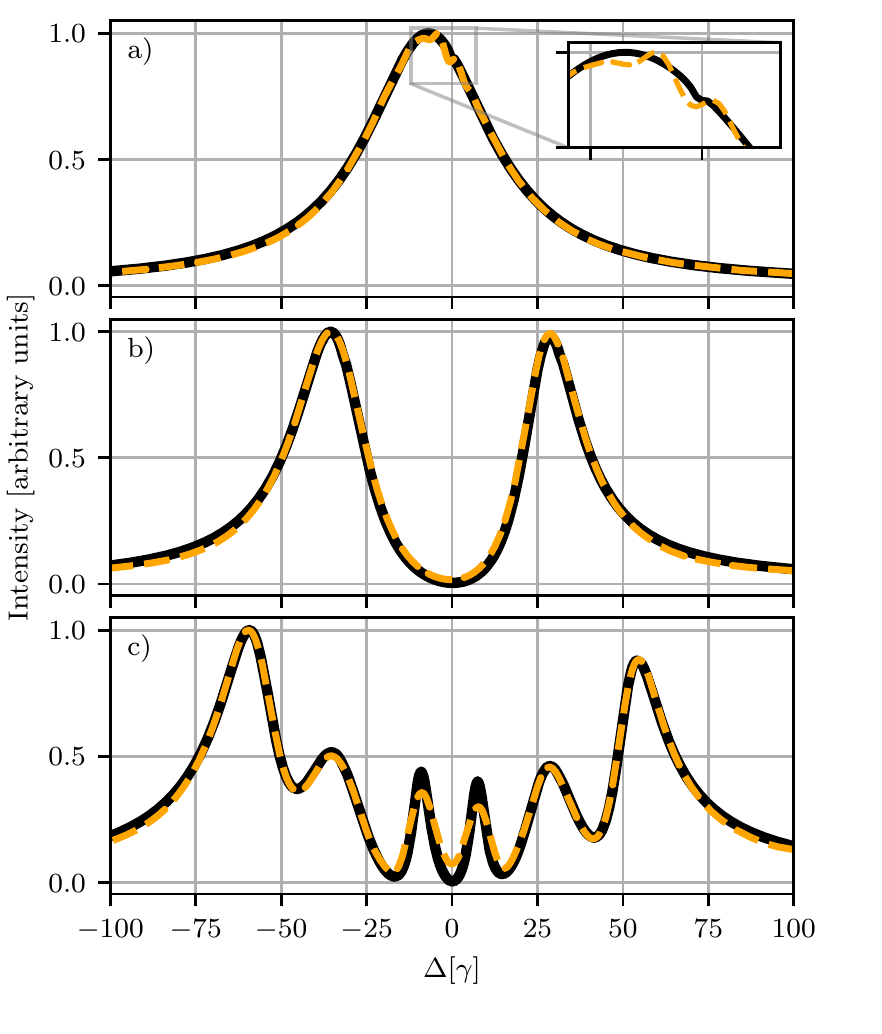}
	\caption{(Color online) Reconstructions of the amplitude of the target's spectral response for the three target settings defined in Fig.~\ref{fig:5_CavityMagnetisation}, as function of the detuning $\Delta = \omega - \omega_0$. The black solid line shows the exact theoretical results, and the yellow dashed line the results from the PHANTASY method. The integration time range is chosen as $t_1 = \unit{15}{ns}$, $t_2 = \unit{ 110}{ns}$ for all three cases.
	}
	\label{fig:5_Oscillation-Drive_reconstruction_good}
\end{figure}

\begin{figure}[t]
	\includegraphics{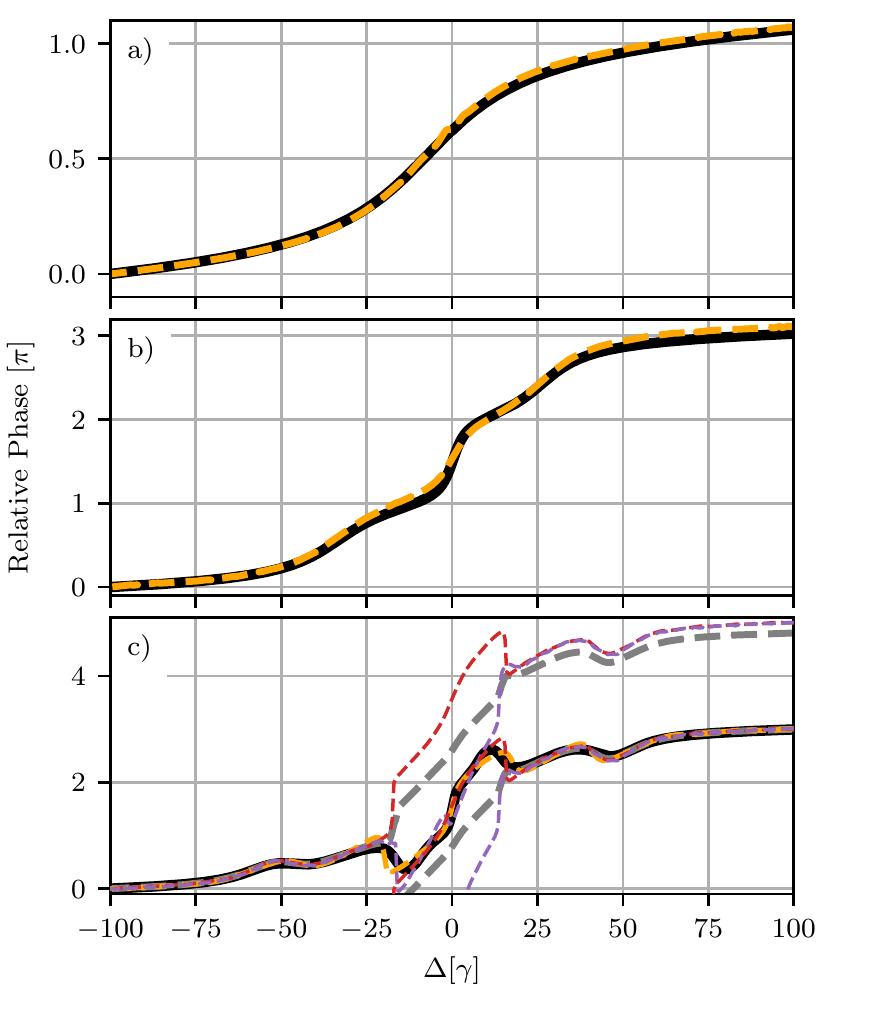}
	\caption{(Color online) Reconstructions of the phase of the target's spectral response for the three target settings defined in Fig.~\ref{fig:5_CavityMagnetisation}, using the PHANTASY method, as function of the detuning $\Delta = \omega - \omega_0$. The black solid line shows the exact theoretical results, and the yellow dashed line the results from the PHANTASY method. In (a,b), the time integration ranges are $t_1 = \unit{15}{ns}$, $t_2 = \unit{130}{ns}$. In (c), the range is $t_1 = \unit{42}{ns}$, $t_2 = \unit{49}{ns}$. The two grey curves further show corresponding results for the range in (a), while the two red (purple) curves show results for the individual time $\unit{80}{ns}$ ($\unit{130}{ns}$). In the latter cases, the two curves each differ only in an overall phase shift, which is chosen such that the curves agree to the reference either at the left or at the right edge of the plot.
	}
	\label{fig:5_Phase Reconstruction_good}
\end{figure}

To perform the simulation, we calculate the intensity registered by the detector as it would be recorded in an experiment using Eq.~\eqref{eq:2_EnergySpectrumOscillatingSL} and the software package pynuss~\cite{pynuss} to calculate the theoretical complex amplitudes of the cavity. Since experimental constraints related to the strong off-resonant component of the incident x-ray pulse typically prevent one from reliably measuring the time-dependent intensity in the first few ns after the excitation, we exclude the first $\unit{15}{ns}$  from the reconstruction. As upper limit for the time range, we use $\unit{192}{ns}$ which corresponds to the bunch separation of the 40-bunch mode at PETRA III (DESY)\cite{DESY_BunchTime}. Both limits are set before any further processing of the data is applied, to simulate the conditions in the experiment.
Based on this truncated data, we then perform the analysis as explained in Sec.~\ref{sec:phantasy}.

\subsubsection{Amplitude of the target's spectral response}
Results for the amplitude of the target spectrum together with corresponding exact reference spectra are shown in Fig.~\ref{fig:5_Oscillation-Drive_reconstruction_good}. For the first configuration without magnetic splitting in (a), a single resonance is observed, as expected. The second configuration (b) features two relatively broad resonances. The third configuration (c) shows all six spectral lines offered by ${}^{57}$Fe. It can be seen that in all cases, the reconstruction agrees very well with the actual target spectra. As expected, the recovery works best for broad spectral features, and becomes worse for more narrow features due to the finite thickness of the reference absorber, as seen for the two central peaks in (c).

However, we note that it is not true in general that very narrow spectral features are not resolved or cannot be captured in the recovery. To illustrate that, the inset of (a) shows a small part around $\Delta = 0$, where the reference spectrum has a very narrow and low-amplitude dip, which is an effect of the finite thickness of the iron layer in the cavity~\cite{CavityDipsDominik}. It can be seen that this tiny spectral feature  is reflected in the recovered spectrum, even though its shape is not accurately captured.

To obtain these spectra, we averaged the data from $t_1 = \unit{15}{ns}$ to $t_2 = \unit{120}{ns}$. This demonstrates that the PHANTASY method indeed enables one to recover the spectra from the early times after the excitation, where the signal intensity is highest. Also, a single time integration range could be used for all three target spectra, illustrating the robustness of the method against the integration range. These aspects will be further analyzed and compared to the Doppler-drive spectroscopy method in Sec.~\ref{sec:Quantitative}.

\begin{figure}
		\includegraphics{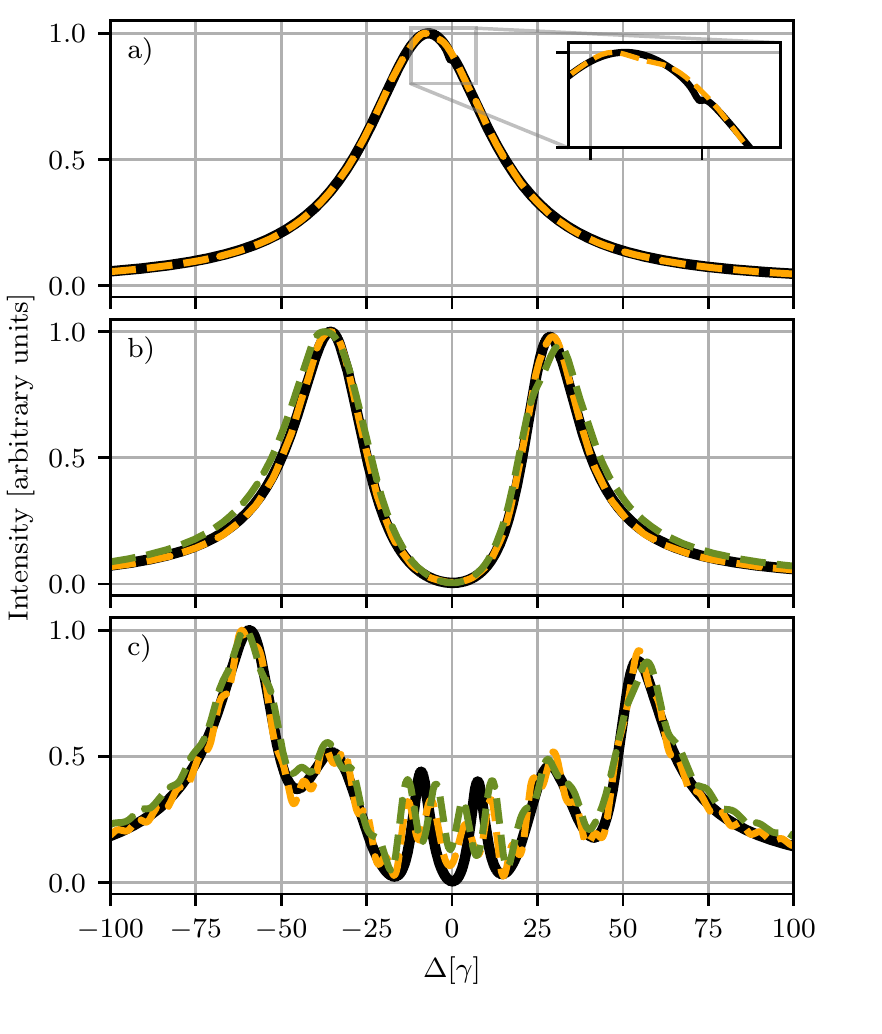}
	\caption{(Color online) Reconstructions of the amplitude of the target's spectral response for the three target settings defined in Fig.~\ref{fig:5_CavityMagnetisation}, using the Doppler-drive method. The black solid line shows the exact theoretical results, and the yellow dashed line the results from the Doppler-drive method. In all three cases, the integration interval $[t_1, t_2]$ and the analyzer thickness $d$ were  optimized separately, in order to achieve best possible results. The parameters are (a) $d = \unit{3}{\mu m}$ and $[117,192]$; (b) $d = \unit{2.5}{\mu m}$, $[175,192]$; c) $d = \unit{3}{\mu m}$ and $[150,192]$. The green dashed curves in (b,c) show corresponding results for the time interval $[117,192]$ of (a) in order to illustrate the stability of the recovery against the integration interval.}
	\label{fig:5_Doppler-Drive_TwoTimes}
\end{figure}

\subsubsection{Phase of the target's spectral response}
Figure~\ref{fig:5_Phase Reconstruction_good} shows the the reconstruction of the spectral phase for the three cases in Fig.~\ref{fig:5_Oscillation-Drive_reconstruction_good}. In all cases, the  overall phase therefore is set to zero at the left boundary of the plot. Similar to the reconstruction of the spectral intensity, in case of setup (a) and (b), the reconstruction of the phase works very well, and is robust against variations in the time-integration range. The results in (a,b) are obtained by integrating from $t_1 = \unit{15}{ns}$ to $t_2 = \unit{130}{ns}$, again showing that the method allows one to use the part of the data with highest intensity. The recovery of the more complicated spectrum in (c) is also good, but less accurate than (a) and (b) especially at $\Delta \approx \pm 10 \gamma$, close to the two resonances with smallest spectral width, where the phase has local extrema and changes rapidly with $\Delta$.

We further found that the reconstruction of the phase is less robust against variations in the time integration range than the reconstruction of the amplitude. The result shown in (c) was obtained by integrating from $t_1 = \unit{43}{ns}$, $t_2 = \unit{53}{ns}$. To illustrate the impact of the integration range, we also show the phase recovered using the same integration range as in (a) and (b) as the two gray dashed lines in Fig.~\ref{fig:5_Phase Reconstruction_good}(c), as well as results for the times $\unit{80}{ns}$ and $\unit{130}{ns}$ individually. In each of the three latter cases, the results are shown twice in the figure. One of the curves is plotted with an overall offset agreeing to the reference at $\Delta = -100\gamma$, whereas the other one has the overall phase fixed to the reference value at $\Delta = + 100\gamma$.  We find that all reconstructions work well and agree to each other except for the region close to the narrow resonances around $\Delta \approx \pm 10 \gamma$. It is important to note that each phase value as function of $\Delta$ is recovered independently. Therefore, even if there are parts in the spectrum which are problematic, e.g., due to narrow spectral structures, it can be expected that the recovery of the other parts is stable against variations in the integration intervals. This feature also allows one to choose suitable integration ranges, and to identify problematic spectral regions, by comparing recoveries for different integration ranges, thereby identifying stable time integration ranges. Furthermore, one can use large time intervals for simple parts of the spectrum in order to make use of most of the measurement statistics, and only restrict the problematic areas of the spectrum to smaller integration ranges to increase the accuracy of the reconstruction.

\subsection{\label{sec:Quantitative}Quantitative analysis of the PHANTASY method}

\begin{figure}[t]
	\includegraphics{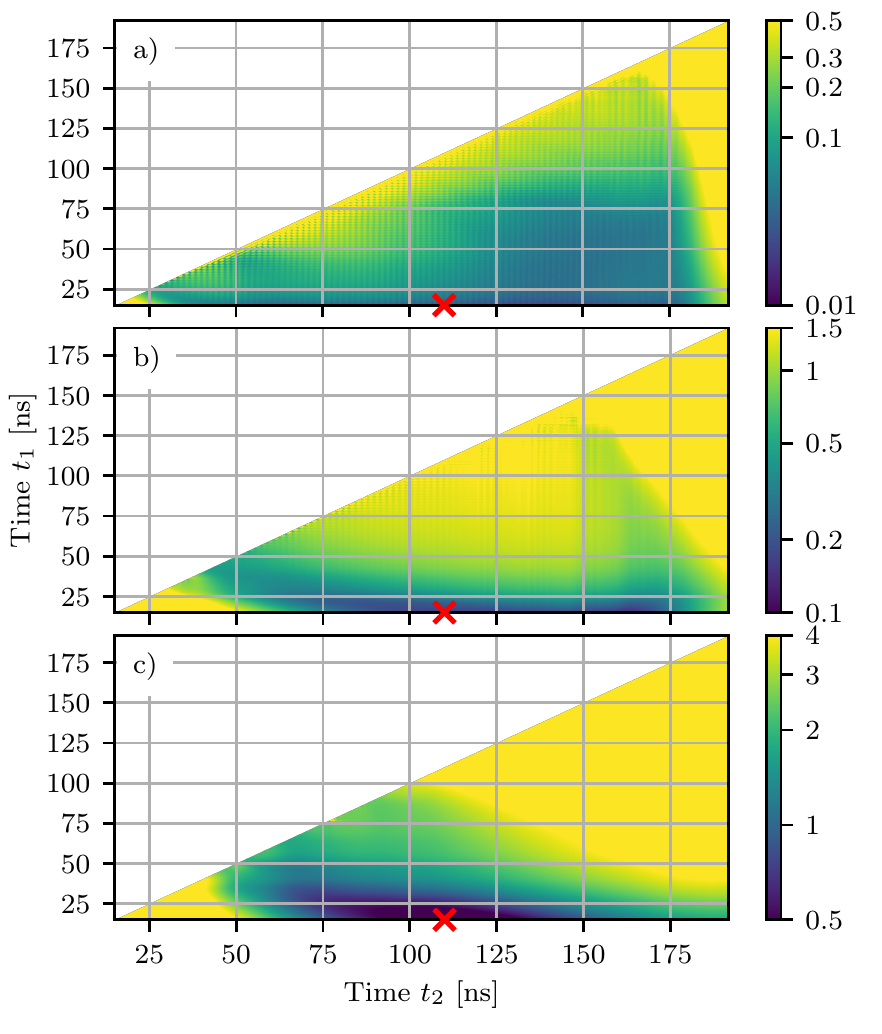}
	\caption{(Color online) Performance of the spectral reconstruction using the PHANTASY method as function of the time-integration range $[t_1,t_2]$. The color encodes the deviation of the reconstructed spectrum from the exact reference calculation. The red crosses indicate the integration ranges used in Fig.~\ref{fig:5_Oscillation-Drive_reconstruction_good}.}
	\label{fig:5_Time_dependency_of_Phantasy}
\end{figure}

\begin{figure}[t]
	\includegraphics{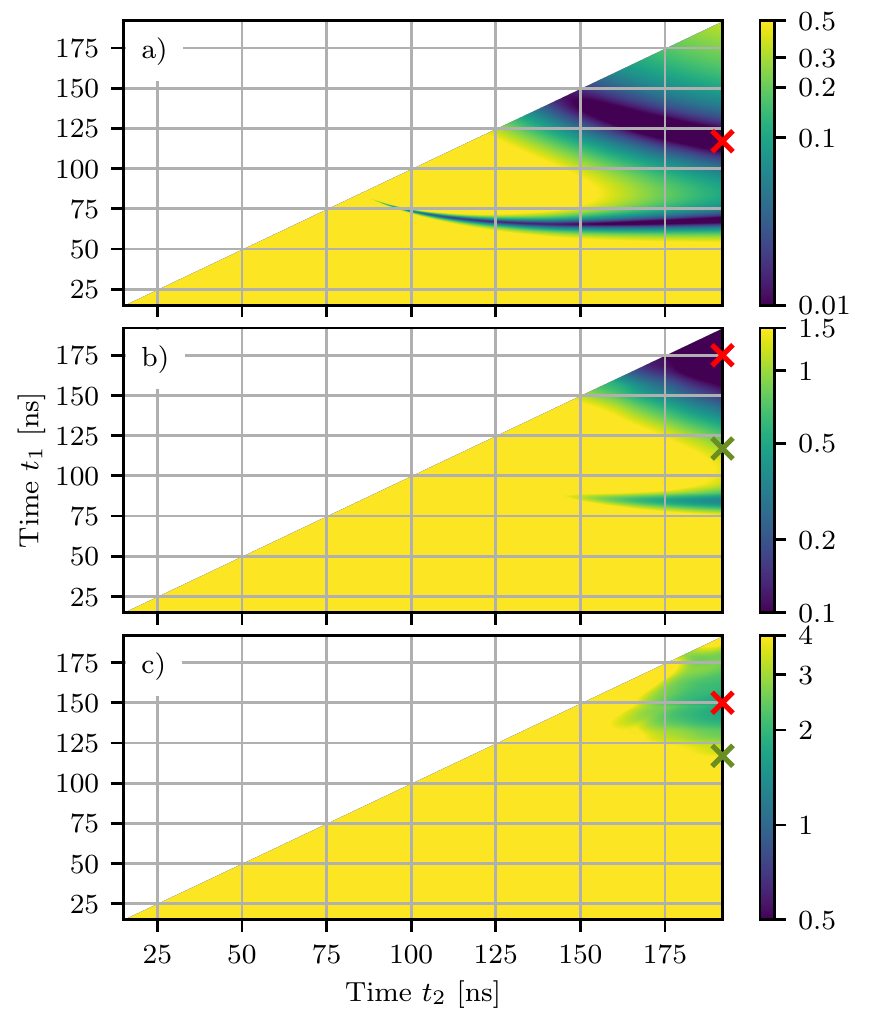}
	\caption{(Color online) Performance of the spectral reconstruction using the Doppler-drive method as function of the time-integration range $[t_1,t_2]$. The color encodes the deviation of the reconstructed spectrum from the exact reference calculation. The red and green crosses indicate the integration ranges used in Fig.~\ref{fig:5_Doppler-Drive_TwoTimes}. }
	\label{fig:5_Time_dependency_of_DopplerDrive}
\end{figure}

\subsubsection{Doppler-Drive spectroscopy method as a reference}

A well-established method to measure spectra using spectrally broad x-ray pulses is the ``Doppler-drive method''~\cite{LTI-LambShift-Roehlberger,Rohlsberger2012,LateTimeIntegrationKilian}. It also assumes the ''Sensing head approximation``, but the analyzer foil is scanned in energy using a Doppler-drive, which moves the analyzer with a constant velocity relative to the target, in contrast to the oscillatory motion of the PHANTASY method. Therefore, no phase $\varphi_0$ appears in the analysis. Also, in the Doppler-drive method, no filters are applied to the recorded data. Instead, the spectrum usually is recovered by integrating over late times only, at which the unwanted contributions to the recorded intensity such as $I_{\textrm{T}}(t)$ ideally have already decayed away.

In order to obtain reference spectra for the comparison with the results of the PHANTASY method, we also simulated the Doppler-drive spectroscopy method for the three cavity  settings in Fig.~\ref{fig:5_CavityMagnetisation}. Since the quality of the recovery depends on the integration range and the analyzer thickness, we optimized $t_1, t_2$, and the thickness for each of the Doppler-drive spectra separately, to obtain the best possible results.

Results are shown in Fig.~\ref{fig:5_Doppler-Drive_TwoTimes}. In (a) and (b), the recovery works well. In (c), the recovery is less good, also connected to the narrow resonances around $\Delta \approx \pm 10$, like in the PHANTASY method. However, the PHANTASY method is able to reconstruct the number of peaks and dips correctly, while the Doppler-drive generates spurious spectral splittings and small residual oscillations across the entire spectrum. Furthermore, to achieve these best results, the integration ranges  $[t_1, t_2]$ had to be chosen in the late time range, where the experimental count rates are low. The best integration ranges are different in the three cases, namely, $[117,192]$ for (a), $[175,192]$ for (b), and $[150,192]$ for (c). Further, the analyzer thickness is $d = \unit{3}{\mu m}$ in (a,c), and $d = \unit{2.5}{\mu m}$ in (b).
To illustrate the impact of the integration range, panels (b,c) also show corresponding results with the integration range of (a) as dashed green curves. It can be seen that in (b), the overall form of the spectrum is still recovered, while quantitatively the  shape is not precisely reproduced. In (c), the number and the shape of the resonances depends on the integration range. Importantly, (b) and (c) show that the deviations between the different integration ranges also appear in the less problematic  areas away from the center of the spectrum. The inset in (a) further shows that the narrow tiny structure in the exact spectrum is not well-recovered by the Doppler drive method.

These results already demonstrate  key advantages of the PHANTASY method over the Doppler-drive method: It allows to recover the spectrum from earlier times than the established Doppler-drive method, such that higher experimental count rates can be included in the analysis. Second, it is more robust against variations in the time integration range. Third, it provides access to the phase of the target response, which is inaccessible in the Doppler-drive method. In the following section, we will analyze the impact of the integration time range in the two methods in more detail.

\subsubsection{\label{sec:comparison}Comparison of suitable integration ranges in the two methods}

In this Section, we analyze the impact of the time-integration range on the performance of the PHANTASY method and the Doppler-drive method in more detail. To this end, we systematically perform the spectral recovery with the two methods as function of the integration ranges $t_1$ and $t_2$. To evaluate the quality of the recovery, we sum the squared difference between the recovered spectra and the corresponding exact theoretical calculation over the recovered spectral range.

\begin{figure}[t]
	\includegraphics{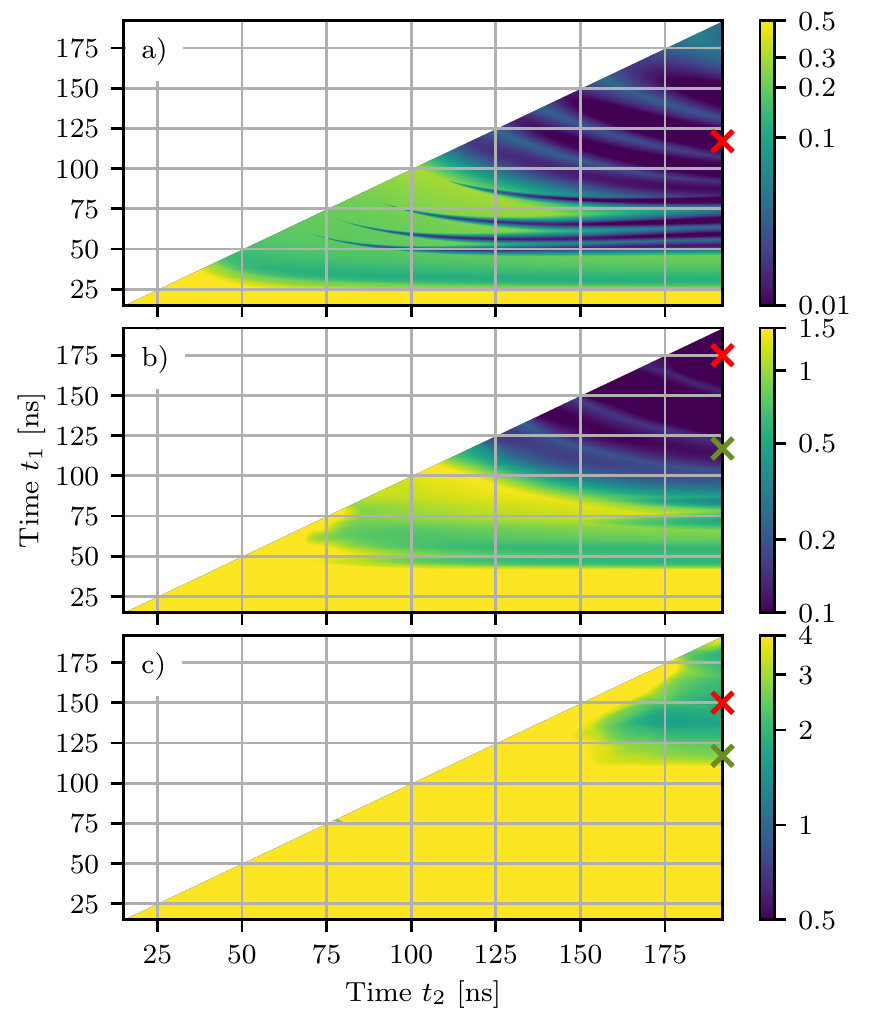}
	\caption{(Color online) Optimum performance of the spectral reconstruction using the Doppler-drive method as function of the time-integration range $[t_1,t_2]$. In comparison to Fig.~\ref{fig:5_Time_dependency_of_DopplerDrive},
	here, the thickness of the analyzer is not kept fixed, but optimized for each pair $t_1,t_2$ separtely, thereby simulating optimum conditions for the Doppler-drive approach. Nevertheless, as compared to the results of the PHANTASY method in Fig.~\ref{fig:5_Time_dependency_of_Phantasy}, integration over early times with high signal intensity remains unfavorable even in this optimum case. }
	\label{fig:opt-b}
\end{figure}

Results for this analysis are shown in Fig.~\ref{fig:5_Time_dependency_of_Phantasy} for the PHANTASY method, and in Fig.~\ref{fig:5_Time_dependency_of_DopplerDrive} for the Doppler-drive method. These plots shows the quality of the reconstruction as function of $t_1$ and $t_2$, where the color encodes the difference between the recovered spectrum and the true spectrum. Dark blue colors indicate good recovery, whereas light yellow colors indicate that the recovery is not fully reliable. White areas appear since $t_2$ must be larger than $t_1$. In both figures, the panels (a,b,c) again correspond to the three target settings defined in Fig.~\ref{fig:5_CavityMagnetisation}. It can be seen that the PHANTASY method works well over a significantly broader range of integration times. Most importantly, in all three cases, best results are achieved if a large integration range is chosen, with a start time $t_1$ chosen close to the lower limit $15$~ns, and the upper limit well beyond $100$~ns. This allows one to include the majority part of the experimental counts into the analysis. In contrast, the Doppler-drive method only works at late integration times, such that only a small fraction of the experimental counts can be used. Furthermore, the PHANTASY method features an area of integration ranges (indicated by the red crosses that show the integration ranges used in Fig.~\ref{fig:5_Oscillation-Drive_reconstruction_good}) which works very well for all three configurations. In contrast, no such common integration range exists for the Doppler method. We note that these general observations  persist even if in the case of the Doppler-drive method, the analyzer thickness is optimized for each integration range $[t_1,t_2]$  separately, thereby simulating optimum conditions for the Doppler-drive approach. The corresponding results are shown in Fig.~\ref{fig:opt-b}. In real experiments, such an optimization usually is not possible, due to a lack of suitable analyzer foils, and since the optimum thickness usually cannot easily be determined {\it a priori}.

We thus conclude that the PHANTASY method is significantly more stable with respect to the choice of the integration range, and in particular allows one to perform the recovery using the early times, where the time-dependent intensity is highest. This allows one to include most of the experimentally recorded signal into the spectral recovery. In contrast, the Doppler-drive method operates at late integration times, restricting the method to only  small parts of the exponentially decaying time-dependent intensity.

\section{\label{sec:summary}Discussion and summary}
In this paper, we introduced the PHANTASY method, which allows for phase-sensitive spectroscopy on nuclear resonances in the hard x-ray regime. Like previous methods, it uses a spectrally narrow resonance absorber in order to introduce spectral information on the relevant energy scales of the nuclei to the measured data. But in contrast to previous spectroscopy methods, in the PHANTASY method, the analyzer performs rapid oscillatory motions along the direction of the x-ray beam. These oscillations lead to the emergence of sidebands in the analyzer response, at multiples of the oscillation frequency $\omega_p$. By tuning this frequency, one of the sidebands is scanned across the spectral response of the target.
The key advantage of the oscillatory motion is the availability of the motional phase $\varphi_0$ at the time of arrival of the x-ray pulse as an additional degree of freedom.

As in other M\"ossbauer spectroscopy approaches, the total detection signal comprises desired components, and spurious background components, which usually cannot be separated from each other  in a straightforward way. We  showed that their different dependencies on $\varphi_0$ allow one to introduce a powerful filter method to separate the total detected intensity into different parts {\it a posteriori}, throughout the data analysis. The method is based on a Fourier-transformation with respect to $\varphi_0$, and a subsequent filtering in the Fourier space.

For the PHANTASY scheme, this $\varphi_0$-filtering enables us to remove those contributions from the measured intensity which arise from photons which scattered only on the target, but not on the analyzer. These do not provide spectroscopy information and therefore form an unwanted background. Note that this filtering is possible even though the desired components and the unwanted background usually are indistinguishable since they overlap in time.

For the PHANTSY method, we augment the $\varphi_0$-filter by a second filter, based on the Fourier transform between time and frequency space. This $t$-filter becomes possible, since after a suitable $\varphi_0$-filtering, the different components contributing to the remaining signal are well-separated in frequency space.
After the two filters, the remaining detection signal is equivalent to that which could have been recorded using a single-sideband analyzer, while retaining the $\varphi_0$-dependence.

In a numerical simulation of the PHANTASY method, we showed that the filtered data enables one to reliably recover the desired amplitude and the phase of the spectral response of the target. A detailed comparison to the established ``Doppler-drive'' method confirmed the key advantages of the PHANTASY method predicted from the analytical results. First, the PHANTASY method is capable of recovering the target spectrum from the detection signal at early times. In contrast, the Doppler-drive method operates at times late after the arrival of the x-ray pulse, where most of the excitation has already decayed. Therefore, a signficantly higher part of the detection signal can be used for the spectral analysis in PHANTASY. Second, we found that the recovery is more stable against variations in the time integration range and the analyzer thickness than the Doppler-drive method. Third, PHANTASY also provides access to the phase of the spectral response, while the Doppler-drive method is restricted to the amplitude only.
For this comparison, we used realistic spectra of different complexity, obtained from thin-film cavities containing nuclei as targets. For these example, we also found that the PHANTASY method is capable of better resolving spectrally more narrow structures than the Doppler-drive method.

Regarding an experimental implementation, we note that the oscillatory motion has the advantage that it is comparably easy to implement and characterize experimentally. However, the $\varphi_0$- and the $t$-filters are required to select the response of a single oscillatory sideband from the total detection signal, to perform the recovery of the target spectrum. As an alternative, we proposed a second motion, which directly leads to a single-sideband response with phase information. This alleviates the need for the filtering, but the motion incurs step-wise jumps, which are more challenging to implement and  characterize experimentally.

Regarding the data analysis, the recovery of the complex target response using PHANTASY discussed here provides a number of avenues to further improve the spectroscopy. For example, the complex response can be verified and refined using self-consistency checks between the recovered spectrum and the measured data, e.g., by calculating the total detection signal expected for the recovered target response and comparing it to the measured data.
Also, since the reconstructed quantity $a(t, \omega_p)$ also contains the complex phase $\varphi_T(t)$ of the target in the time domain, one can extend the evaluation such that this phase is reconstructed as well. Since this allows one to independently determine the time- and the frequency-domain phases of the target response, they can be verified against each other using a Fourier transform. Using an iterative algorithm, this cycle can be performed several times until the measured data and the recovered data in the time- and frequency domains are self consistent.
We further suggest to perform an analysis similar as  in Figs.~\ref{fig:5_Time_dependency_of_Phantasy} and~\ref{fig:5_Time_dependency_of_DopplerDrive} to optimize the averaging interval $[t_1, t_2]$. Since the theoretical reference obviously is not known, it can be replaced by the recovered response for a particular interval. Alternatively, the plot can be modified to display the difference of the recovered spectrum to the neighboring intervals. This way, stable $[t_1, t_2]$ intervals covering as much of the measured data as possible can be determined, and the quality of the reconstruction can be judged.

Finally, we note that the PHANTASY method crucially relies on an event-based detection method, which allows one to analyze the data  after the experiment with respect to an arbitrary combination of the experimental parameters. For instance, this capability is the key requirement for the  $\varphi_0$- and $t$-filters proposed here, and we expect that these methods to separate otherwise indistinguishable signal contributions  will find further applications beyond the PHANTASY method. Most importantly, the multi-dimensional measurement as function of various  parameters such as time, detuning, oscillation frequency and amplitude allows one to perform much more stringent comparisons to theoretical predictions than the established one-dimensional spectra as function of only a single variable. Using such multi-dimensional fits to the data, ultimately, we  envision a direct model-independent reconstruction of amplitude and phase of arbitrary target spectra. We have already demonstrated a similar approach to recover the precise piezo motion of a reference absorber in a model-independent way~\cite{2dSpectra}, or to recover the nuclear quantum dynamics coherently controlled  by a suitably shaped x-ray pulse~\cite{coherentcontrol}. Promoting this approach to a full spectroscopy technique has the additional advantage that the sensing head approximation is no longer required, thereby providing a perspective for the recovery of  target spectra independent of source- and analyzer broadenings.

\section*{Acknowledgments}
We thank Dominik Lentrodt and Kilian P. Heeg for valuable discussions. 

\bibliography{Literatur}

\end{document}